\pdfoutput=1
\documentclass[11pt]{article}

\usepackage[preprint]{acl}
\usepackage{multirow}
\usepackage{graphicx}
\usepackage{amssymb}
\usepackage[table]{xcolor}
\usepackage{array}

\makeatletter
\def\thickhline{%
  \noalign{\ifnum0=`}\fi\hrule \@height 1.1pt \futurelet\reserved@a\@xhline}
\makeatother
\usepackage{times}
\usepackage{latexsym}

\usepackage[T1]{fontenc}
\usepackage[utf8]{inputenc}

\usepackage{microtype}

\usepackage{inconsolata}
\usepackage{twemojis}
\definecolor{speechrow}{RGB}{255,235,210}
\definecolor{audiorow}{RGB}{222,238,255}
\definecolor{musicrow}{RGB}{225,245,225}
\usepackage{algorithm}
\usepackage{algorithmic}
\usepackage{multirow}
\usepackage{booktabs}
\usepackage{amsmath}
\usepackage{amssymb}
\usepackage{graphicx}
\usepackage{bbding}

\usepackage{newfloat}
\usepackage{listings}

\title{Audio Editing in the Era of Foundation Models: A Survey}

\author{
    Changhao Pan$^{1}$\thanks{$\,$ Equal Contribution.},
    Yifei Fan$^{1}$\footnotemark[1],
    Fan Zhuo$^{1}$\footnotemark[1],
    Yifu Chen$^{1}$, Wenxiang Guo$^{1}$,
    \\
    \textbf{
    Yu Zhang$^{2}$, Ruiqi Li$^{2}$, Zhiyuan Zhu$^{1}$, Rui Yang$^{1}$, Shengpeng Ji$^{3}$,
    } \\
    \textbf{
    Chenyuhao Wen$^{1}$, Jiayang Xu$^{1}$, Ke Lei$^{1}$, Xiaoda Yang$^{1}$, Jingyu Lu$^{1}$, Zhou Zhao$^{1}$\thanks{$\,$ Corresponding Author.}
    } \\
    $^{1}$Zhejiang University \quad $^{2}$Bytedance \quad $^{3}$Hunyuan Team, Tencent \\
    \texttt{panch@zju.edu.cn, yifei1.23@intl.zju.edu.cn, zhaozhou@zju.edu.cn}
}

\begin{document}
\maketitle
\begin{abstract}
% 第一段，背景内容：任务是什么->发展现状->为什么需要survey
Audio editing aims to modify a given synthetic or real-world audio signal to satisfy specific user needs. 
As a promising yet challenging direction in AIGC, it has attracted increasing attention. Recent advances in audio generation have made powerful generative models central to modern audio editing systems. 
This rapid progress has created a growing need to organize emerging tasks, methods, and resources into a coherent view.
% 第二段，综述内容，写的自然连贯一些
In this survey, we provide a comprehensive review of audio editing in the era of foundation models. We first present a unified taxonomy of existing editing tasks and then summarize the major foundation-model paradigms that support modern audio editing, covering representative approaches from both training-based and training-free perspectives. We further discuss related resources, including datasets, evaluation protocols, and data construction tools. 
Finally, we identify open challenges in this field and outline promising directions for future research. 
The accompanying repository is released at \url{https://github.com/DaViD-Pigeon/AudioEditSurvey}.

\end{abstract}

% intro + abs 5/3
\section{Introduction}

Nowadays, in the territory of AI-generated content (AIGC)~\cite{deepseek2026deepseekv4,seedance2026seedance,comanici2025gemini}, audio editing has emerged as a crucial technique for practical applications in commerce, entertainment, and scientific research~\cite{du2025cosyvoice,lin2025omnihuman}. Unlike audio generation, which creates new audio from scratch, audio editing aims to modify a given recording according to user intent, such as changing its content, background, style, speaker identity, or acoustic conditions. These edits may range from fine-grained local modifications to global switches over the entire timeline, while preserving task-irrelevant acoustic content and temporal coherence.

Early audio editing workflows were heavily dependent on digital audio workstations\footnote{\url{https://www.adobe.com/products/audition.html}} and signal-processing tools\footnote{\url{https://www.steinberg.net/cubase}}, requiring substantial manual effort and domain expertise. With the rapid progress of deep generative models~\cite{liu2022flow,ho2020denoising,goodfellow2014generative}, this paradigm is shifting toward data-driven and foundation-model-based approaches. Recent advances in diffusion models~\cite{rombach2022high,peebles2023scalable,esser2024scaling} and audio-language models~\cite{kong2024audioflamingo,kreuk2022audiogen} have significantly broadened the boundary of audio editing, enabling more flexible, semantic, and instruction-driven manipulation. 

Despite these rapid advances, foundation-model-based audio editing has not been systematically reviewed. Existing surveys on the audio modality mainly focus on audio-language models~\cite{latif2023sparks,su2025audio}, spoken language models~\cite{ji2024wavchat}, and general audio generation~\cite{xie2025towards,bovzic2024survey}. While some of them briefly mention editing-related tasks~\cite{ma2024foundation}, audio editing is often discussed only as a minor component of broader audio generation or manipulation frameworks. 
Thus, to address this gap, we conduct this survey to offer a comprehensive analysis specifically focused on audio editing in the era of foundation models.

\begin{figure*}[t]
    \centering
    \includegraphics[width=\textwidth]{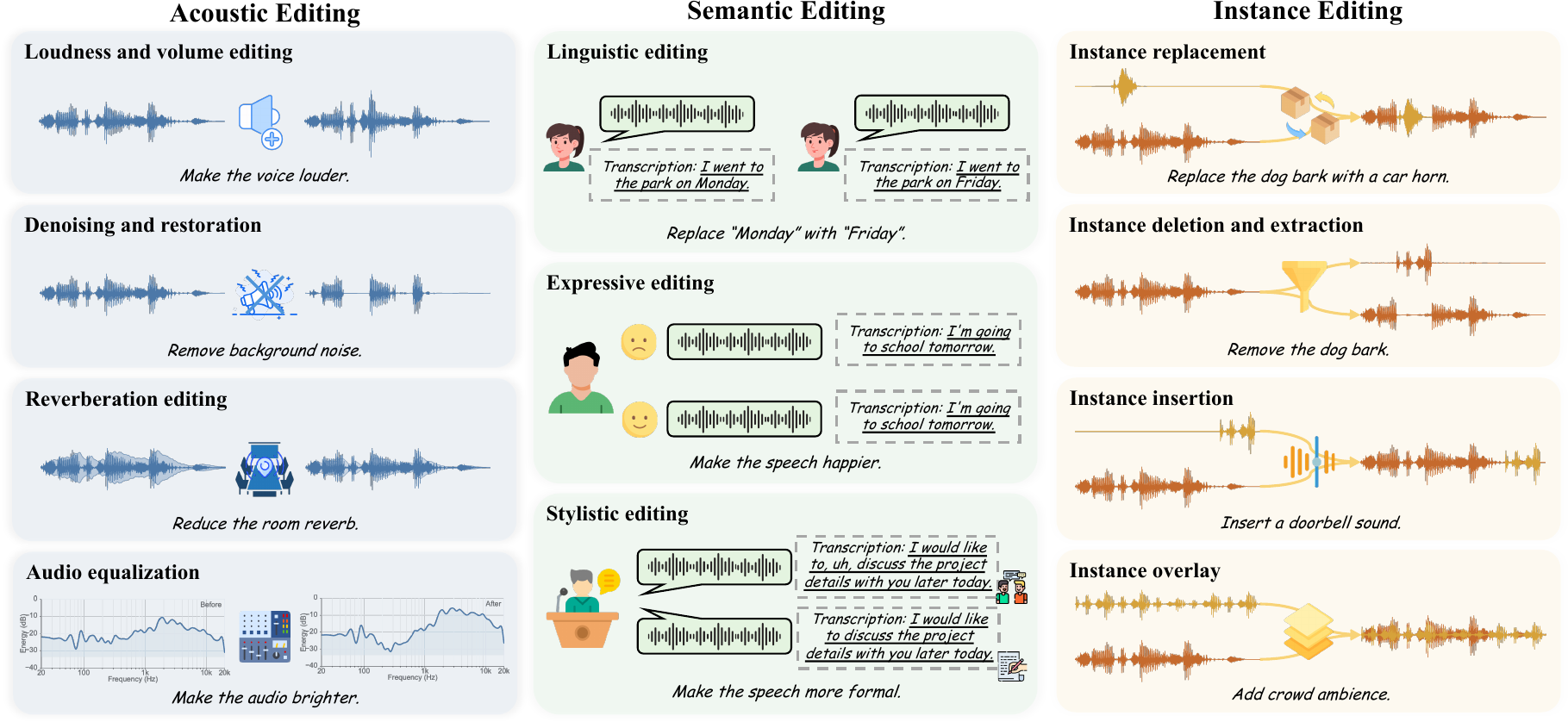}
    \caption{
Taxonomy of audio editing tasks.
    }
    \label{fig:task}
    \vspace{-0.5em}
\end{figure*}

In particular, this survey offers a systematic analysis of audio editing from four perspectives: editing tasks, foundation-model architectures, learning paradigms, and resources. First, in Section~\ref{sec:task}, we review representative studies in recent audio editing research and propose a task taxonomy based on input conditions and expected editing outcomes. This taxonomy divides audio editing into three major categories, namely acoustic editing, semantic editing, and instance editing, covering twelve fine-grained editing tasks. It highlights how diverse editing goals impose distinct requirements on controllability, temporal localization, and content preservation.
Second, we examine the mainstream foundation-model architectures adopted for audio editing, and analyze their suitability for different editing scenarios in Section~\ref{sec:arch}. Then, we organize existing methods according to their learning strategies, distinguishing between training-based and training-free approaches, and further categorize them by their core technical mechanisms. In addition, we provide a comprehensive summary of resources for audio editing, including public datasets, data construction tools, evaluation benchmarks, and metrics. We also discuss key challenges and future directions in this rapidly evolving field.

In summary, this survey aims to systematically classify and evaluate a large body of research in the field of audio editing under foundation models. We hope that our work can provide a comprehensive resource that not only synthesizes current findings but also guides future research directions in this rapidly evolving field.

% 先这么放着，到时候调整的时候丢到附录去（TODO: 检查intro abs 和 taxnomy是不是需要修正）

% 具体任务丢到附录，压缩的到每个subsection都是分类概述（1段）+具体任务（1段），每个任务一句话

\section{Taxonomy} \label{sec:task}

To systematically organize the audio editing in the foundational-model era, a taxonomy that is defined by the primary object itself is needed to answer one particular question: \textit{which layer of the information of the audio is edited by the model?} As a result, we organize current audio editing tasks into the following three representative categories. Refer to Appendix~\ref{app_sec:task} for task details.

\noindent\textbf{Acoustic editing} refers to transformations of signal-level attributes that affect auditory perception without changing the underlying semantic content, source identity, or temporal organization. It treats loudness, noise, reverberation, bandwidth, and spectral balance as controllable acoustic conditions. Representative tasks include \textit{loudness and volume editing} for adjusting component salience~\cite{jiang2024listen,choi2021amss,liang2024wavcraft}; \textit{denoising and restoration} for removing recording or transmission degradations~\cite{liu2022voicefixer,fu2021metricgan+}; \textit{reverberation editing} for modifying room response and decay patterns~\cite{chen2022visual,bahrman2024speech}; and \textit{audio equalization} for reshaping spectral coloration~\cite{venkatesh2022word}.

\noindent\textbf{Semantic editing} operates on high-level, interpretable information conveyed by audio, such as linguistic content, expressive intent, affective state, or stylistic semantics, while preserving non-target factors including speaker identity, source characteristics, acoustic context, and temporal coherence. Across speech, music, and general audio, it changes what the signal communicates or how it is perceived by listeners. \textit{Linguistic editing} revises explicit language-bearing content within an existing recording~\cite{tan2021editspeech,zhang2022editsinger}. \textit{Expressive editing} modifies the manner in which audio content is delivered or emotionally perceived~\cite{wang2024emotion,kim2025fillerspeech}. \textit{Stylistic editing} changes the higher-level presentation style that shapes the overall character of speech, music, or sound events~\cite{zhang2024musicmagus,anastassiou2024voiceshop,jin2023voice}.

\noindent\textbf{Instance editing} focuses on source-level entities in an audio scene, such as speakers, sound events, instruments, vocal tracks, or background components. Instead of primarily changing acoustic conditions or communicated meaning, it modifies the presence, identity, or source-to-mixture relation of selected entities while preserving non-target content, scene coherence, and temporal structure. \textit{Instance replacement} substitutes a target entity with another source~\cite{wang2021vqmivc,qian2019autovc,xu2024prompt,han2023instructme,fu2025object}. \textit{Instance deletion and extraction} suppress or isolate selected sources from a mixture~\cite{ellis2025recomposer,liu2024separate,vzmolikova2019speakerbeam,defossez2021hybrid}. \textit{Instance insertion} introduces new entities into an existing recording~\cite{wang2023audit,jin2017voco,rouard2025musicgen,peng2024voicecraft}. \textit{Instance overlay} layers or emphasizes sources to enrich, rebalance, or remix an audio scene~\cite{ellis2025recomposer,strano2025stage}.

\section{Foundation Models for Audio Editing}\label{sec:arch}

% \subsection{GAN-based generative paradigm }
% In the early exploratory line, GAN method was first used for audio editing. The earliest starting point is A GAN Speech Inpainting Model for Audio Editing Software\cite{todo}, which conducts inpainting in assigned time-frequency regions through GAN model, while Emotion Selectable End-to-End Text-Based Speech Editing\cite{todo} can not only execute text-based speech editing that targets on content, but also alters emotional components in the edited areas through generative adversarial framework.

\subsection{Early Neural Editing Models}

Before the foundation-model era, early neural audio editing methods explored task-specific generative models for local reconstruction and attribute control.
For example, GAN-based speech inpainting reconstructs missing or corrupted time-frequency regions for localized repair~\cite{zhao2023gan}, while adversarial text-based speech editing modifies target regions in linguistic content and emotional attributes~\cite{wang2024emotion}.
Although limited in scale and generalization, these methods introduced core ideas such as local generation, context-aware reconstruction, boundary consistency, and attribute-conditioned editing.

\subsection{Token-based Codec Language Models}
% 范式定义
Token-based codec language models cast audio editing as conditional generation over discrete audio tokens. After continuous audio is converted into compact discrete token sequences, target regions are edited through autoregressive continuation, infilling, or selective regeneration conditioned on context, prompts, or task controls.

% 介绍一下codec工作的生成模型基础
This paradigm is enabled by recent token-based audio generation models, which establish discrete tokens as a unified interface for modeling speech and music. For example, AudioLM~\cite{borsos2023audiolm} represents speech with semantic and acoustic tokens, while VALL-E~\cite{wang2023neural} applies neural codec language modeling to zero-shot speech synthesis with speaker prompts. MusicGen~\cite{copet2023simple} extends this idea to text-conditioned music generation, and SoundStorm~\cite{borsos2023soundstorm} improves decoding efficiency by parallel iterative token generation.

% 编辑化转变的工作
The explicit temporal structure of token sequences further makes this paradigm suitable for localized editing. VoiceCraft~\cite{peng2024voicecraft} treats a target region within an existing speech-token sequence as an infilling span, enabling zero-shot speech editing without regenerating the full utterance. SpeechX~\cite{wang2024speechx} generalizes token-based editing to multiple speech transformation tasks, using discrete tokens as a shared representation for editing, extraction, and conversion. Later studies improve robustness and selectivity through multi-span editing, background-aware insertion and replacement, hallucination suppression, and stricter preservation of non-target areas~\cite{wang2024maskgct,wang2025ssr,chen2025seamlessedit}. 

% 整体评价和优劣势分析
% Overall, token-based codec language models are well suited for speech content editing, continuation, and speaker-conditioned infilling, as they provide compact sequence representations compatible with language model architectures. However, they may suffer from autoregressive latency, error accumulation, and quality degradation when editing music or general audio with broad spectral dynamics, partly due to the high compression of discrete tokens.

% TODO: 
% 1. Caption 幻觉修改 
% 2. 工作缩写调整，很多直接用的标题看起来就不对； 
% 3. 确认一下text这个condition是否合理 要不要改成inst？（和training/training-free的condtion对应）（已经完成，决定不改）
% 4. 边界收窄之后需要重新打checkmark
\begin{table*}[t]
\centering
\scriptsize
\renewcommand{\arraystretch}{1.12}
\caption{\textbf{Category definitions.}
Row colors indicate domains: \colorbox{speechrow}{speech}, \colorbox{audiorow}{general audio}, and \colorbox{musicrow}{music}.
\textbf{Codec} and \textbf{Diff.} denote codec-based and diffusion-based models, respectively.
\textbf{Text} denotes natural-language instructions, prompts, or transcript guidance; \textbf{Mask} denotes explicit mask or region guidance; \textbf{Ref.} denotes reference-audio conditioning.
Editing categories follow our taxonomy: 
\textbf{Rest} denotes denoising and restoration; 
\textbf{Rev} denotes reverberation editing; 
\textbf{Loud} denotes loudness or volume editing; 
\textbf{EQ} denotes audio equalization; 
\textbf{Ling} denotes linguistic editing; 
\textbf{Expr} denotes expressive editing; 
\textbf{Sty} denotes stylistic editing; 
and \textbf{Rep}, \textbf{Del}, \textbf{Ins}, and \textbf{Ovl} denote instance replacement, deletion or extraction, insertion, and overlay, respectively.}
\label{tab:audio_editing_methods}
\vspace{-2pt}
\resizebox{\textwidth}{!}{
\begin{tabular}{l c c c c c c c c c c c c c}
\toprule

\multicolumn{1}{c}{\multirow{2}{*}{\textbf{\small Method}}}
& \multicolumn{1}{c}{\multirow{2}{*}{\textbf{\small Arch}}}
& \multicolumn{1}{c}{\multirow{2}{*}{\textbf{\small Cond}}}
& \multicolumn{4}{c}{\textbf{\small Acoustic Editing}}
& \multicolumn{3}{c}{\textbf{\small Semantic Editing}}
& \multicolumn{4}{c}{\textbf{\small Instance Editing}} \\[1.5pt]

\cmidrule(lr){4-7}\cmidrule(lr){8-10}\cmidrule(lr){11-14}

\multicolumn{1}{c}{}
& \multicolumn{1}{c}{}
& \multicolumn{1}{c}{}
& \multicolumn{1}{c}{\textbf{\small Rest}}
& \multicolumn{1}{c}{\textbf{\small Rev}}
& \multicolumn{1}{c}{\textbf{\small Loud}}
& \multicolumn{1}{c}{\textbf{\small EQ}}
& \multicolumn{1}{c}{\textbf{\small Ling}}
& \multicolumn{1}{c}{\textbf{\small Expr}}
& \multicolumn{1}{c}{\textbf{\small Sty}}
& \multicolumn{1}{c}{\textbf{\small Rep}}
& \multicolumn{1}{c}{\textbf{\small Del/Ext}}
& \multicolumn{1}{c}{\textbf{\small Ins}}
& \multicolumn{1}{c}{\textbf{\small Ovl}}\\[1pt]
\midrule

\multicolumn{14}{c}{\textit{Training-based Methods}} \\ 
\midrule

\rowcolor{speechrow}
FluentSpeech~\cite{jiang2023fluentspeech} 
& Diff.
& Text, Mask 
&  &  &  &  
& $\checkmark$ & $\checkmark$ &  
&  & $\checkmark$ & $\checkmark$ &  \\

\rowcolor{speechrow}
VoiceCraft~\cite{peng2024voicecraft}
& Codec
& Text 
&  &  &  &  
& $\checkmark$ &  &  
&  &  &  &  \\

\rowcolor{speechrow}
uSee~\cite{yang2024usee}
& Diff.
& Text
& $\checkmark$ & $\checkmark$ & $\checkmark$ &
&  &  &
&  &  & $\checkmark$ & $\checkmark$ \\

\rowcolor{speechrow}
Step-Audio-EditX~\cite{yan2025step}
& Codec
& Text
& $\checkmark$ &  &  &
&  & $\checkmark$ & $\checkmark$
&  &  & $\checkmark$ &  \\

\rowcolor{speechrow}
InstructSpeech~\cite{huang2024instructspeech}
& Codec
& Text
&  &  & $\checkmark$ &  
& $\checkmark$ & $\checkmark$ & $\checkmark$
&  &  &  &  \\

\rowcolor{audiorow}
AUDIT~\cite{wang2023audit}
& Diff.
& Text 
& $\checkmark$ &  &  &  
&  &  &  
& $\checkmark$ & $\checkmark$ & $\checkmark$ & $\checkmark$ \\

\rowcolor{audiorow}
SAO-Instruct~\cite{ungersbock2025sao}
& Diff.
& Text 
& $\checkmark$ &  & $\checkmark$ & $\checkmark$ 
&  & $\checkmark$ & $\checkmark$ 
& $\checkmark$ & $\checkmark$ & $\checkmark$ & $\checkmark$ \\

\rowcolor{audiorow}
Non-Rigid Prompt Edit~\cite{paissan2023audio}
& Diff.
& Text 
& $\checkmark$ &  &  &
& $\checkmark$ &  & $\checkmark$
& $\checkmark$ &  & $\checkmark$ & $\checkmark$ \\

\rowcolor{musicrow}
InstructME~\cite{han2023instructme}
& Diff.
& Text, Ref.
&  &  & $\checkmark$ & $\checkmark$ 
&  &  & $\checkmark$ 
& $\checkmark$ & $\checkmark$ & $\checkmark$ & $\checkmark$ \\

\rowcolor{musicrow}
Instruct-MusicGen~\cite{zhang2024instruct}
& Codec
& Text 
&  &  & $\checkmark$ &  
&  &  & $\checkmark$ 
&  & $\checkmark$ & $\checkmark$ & $\checkmark$ \\

\midrule
\multicolumn{14}{c}{\textit{Training-free Methods}} \\ 
\midrule

\rowcolor{speechrow}
AST~\cite{lv2026ast}
& Diff.
& Text
&  &  &  &
& $\checkmark$ & $\checkmark$ & $\checkmark$
&  &  &  &  \\

\rowcolor{speechrow}
EdiTTS~\cite{tae2021editts}
& Diff.
& Text, Mask
&  &  &  &
& $\checkmark$ & $\checkmark$ &
&  &  &  &  \\

\rowcolor{audiorow}
DDPM Inversion~\cite{manor2024zero}
& Diff.
& Text
&  &  &  &  
&  &  & $\checkmark$ 
& $\checkmark$ & $\checkmark$ &  &  \\

\rowcolor{audiorow}
AudioEditor~\cite{jia2025audioeditor}
& Diff.
& Text
&  &  &  &  
&  &  & $\checkmark$ 
& $\checkmark$ & $\checkmark$ & $\checkmark$ &  \\

\rowcolor{audiorow}
PPAE~\cite{xu2024prompt}
& Diff.
& Text 
&  &  & $\checkmark$ &  
&  &  & $\checkmark$ 
& $\checkmark$ & $\checkmark$ & $\checkmark$ & $\checkmark$ \\

\rowcolor{audiorow}
AudioMorphix~\cite{liang2025audiomorphix}
& Diff.
& Ref., Mask 
&  &  & $\checkmark$ & $\checkmark$ 
&  &  &  
& $\checkmark$ & $\checkmark$ & $\checkmark$ & $\checkmark$ \\

\rowcolor{musicrow}
MelodyFlow~\cite{lan2024high}
& Diff.
& Text
&  &  &  &  
&  &  & $\checkmark$ 
&  &  &  &  \\

\rowcolor{musicrow}
MEDIC~\cite{liu2024medic}
& Diff.
& Text
&  &  &  & $\checkmark$ 
&  & $\checkmark$ & $\checkmark$ 
& $\checkmark$ & $\checkmark$ & $\checkmark$ &  \\

\rowcolor{musicrow}
MusicMagus~\cite{zhang2024musicmagus}
& Diff.
& Text
&  &  &  & $\checkmark$ 
&  & $\checkmark$ & $\checkmark$ 
& $\checkmark$ &  &  &  \\

\rowcolor{musicrow}
MusRec~\cite{boudaghi2025musrec}
& Diff.
& Text
&  &  &  & $\checkmark$ 
&  &  & $\checkmark$ 
& $\checkmark$ &  &  &  \\

\bottomrule
\end{tabular}
}
\vspace{-0.5em}
\end{table*}

\subsection{Diffusion and Flow-Matching Models}
% 范式定义
Diffusion and flow-matching models formulate audio editing as conditional transformation in continuous acoustic spaces, such as mel-spectrograms or audio latents. Instead of infilling discrete tokens, they modify audio through conditional denoising, latent inversion, or continuous flow transformation, making them suitable for high-fidelity reconstruction, region-level refinement, and fine-grained acoustic control in complex scenarios.

% 介绍一下diffusion based的生成模型基础
Diffusion-based audio generation models provide the foundation for this paradigm by establishing iterative denoising as a strong framework for speech, general audio, and music synthesis. Representative examples include WaveGrad~\cite{chen2020wavegrad} for waveform generation, F5TTS~\cite{chen2025f5} and MegaTTS3~\cite{jiang2025megatts} for text-to-speech synthesis, AudioLDM~\cite{liu2024audioldm} for text-to-audio generation, and Mo{\^u}sai~\cite{schneider2023mousai} and Stable Audio~\cite{evans2024fast} for text-to-music generation. These models largely motivate later editing methods that adapt full-sample synthesis to source-conditioned and region-constrained modification.

% 编辑化的工作
The continuous generative process further makes this paradigm suitable for constrained audio modification. Editing-oriented methods adapt denoising or flow transformation by conditioning on source audio, masks, target regions, reference signals, or textual instructions, so that only the intended content is regenerated. FluentSpeech~\cite{jiang2023fluentspeech} and DiffEditor~\cite{chen2024diffeditor} apply diffusion models to localized speech refinement and text-based editing, improving contextual reconstruction and boundary consistency. uSee~\cite{yang2024usee} extends diffusion editing to acoustic control with environmental conditions, while CosyEdit~\cite{chen2026cosyedit} uses flow matching to improve region-specific consistency and alignment. Beyond speech, AUDIT~\cite{wang2023audit}, PPAE~\cite{xu2024prompt}, and AudioMorphix~\cite{liang2025audiomorphix} develop instruction-guided, event-localized, and reference-guided audio editing methods, and MelodyFlow~\cite{lan2024high} further extends continuous latent editing to music through flow matching and latent inversion.

\subsection{Audio Editing Interfaces}

Instructional and multimodal interfaces enable high-level control for foundation-model-based audio editing by converting textual instructions, reference audio, temporal regions, or visual cues into target spans, task embeddings, event locations, speaker references, or preservation constraints.

Representative works enable flexible, user-driven editing through such interfaces. Early efforts translate user requests into task-level or token-level control representations for unified editing, extraction, and conversion~\cite{huang2024instructspeech,wang2024speechx}. Recent studies extend these interfaces to finer-grained and more reliable control, including multi-span editing, background-aware modification, hallucination suppression, and non-target preservation~\cite{wang2025ssr,chen2025seamlessedit,huang2025voicenong,wang2023audit}. These advances support broader scenarios such as sound-effect insertion, object replacement, soundscape remixing, and instructional music editing.

\section{Training-based Approaches}\label{sec:training-based}
\begin{figure*}[t]
    \centering
    \includegraphics[width=\textwidth]{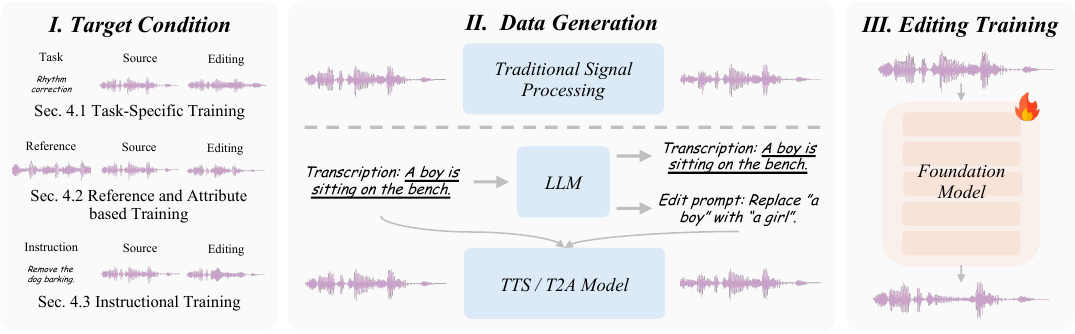}
    \caption{
Overview of training-based audio editing methods.
    }
    \label{fig:training-based}
    \vspace{-0.5em}
\end{figure*}

Training-based approaches learn editing behaviors from paired, pseudo-paired, or instructional supervision. By optimizing editing objectives and preservation constraints, they provide stable control over predefined or user-specified edits. As shown in Figure~\ref{fig:training-based}, we categorize them according to their supervision and conditioning mechanisms.

\subsection{Task-Specific Training}
% 方法范式定义
Task-specific training optimizes audio editing models for predefined functions, such as speech editing, prosody correction, source separation, and music stem separation.
% 方法介绍
It typically relies on paired data with structured task-dependent conditions, including transcripts, masked regions, alignment information, language queries, or source labels.

% Speech Edit
In speech editing, early methods focus on localized reconstruction and preservation, using partial inference, bidirectional fusion, or alignment-aware acoustic-text representations to reduce artifacts in edited and unmodified regions~\cite{tan2021editspeech,bai20223}. 
Later studies further address context and boundary consistency through in-context prosody modeling and attention-based boundary stitching~\cite{yu2023cross,alexos2024attentionstitch}. 

% 其他领域
Beyond speech, task-specific training supports instance-level manipulation by isolating queried sounds or musical sources from mixtures through language-conditioned or dedicated separation models~\cite{liu2022separate,liu2024separate,rouard2023hybrid}.

% 总结一下适用场景和优缺点
This paradigm is reliable for well-defined edits, but its flexibility is often limited by the task and data distributions used in training.

\subsection{Reference and Attribute based Training}

% 训练范式定义和依赖说明
Reference and attribute-based training specifies editing directions through external conditions, such as reference audio, style examples, or attribute labels. It typically relies on paired, weakly paired, or contrastive data linking input audio to target conditions. 
% 和Task-Specific的区别
Compared with task-specific training, its conditions are more flexible and broadly applicable, covering references, categorical labels, descriptive attributes, or relative attribute differences.

\noindent\textbf{Reference-based Training}
uses external audio examples to specify target styles, such as production characteristics, speaker timbre, accent, or emotion. 
% 音效相关
In audio production, related methods, such as DeepAFx-ST, transfer reference production and mixing characteristics through effect-parameter prediction or style feature extraction~\cite{steinmetz2022style,koo2023music}. 
% Speech Edit
In speech editing, reference speech guides timbre transfer and real-time voice style conversion~\cite{li2024sef,liu2026stylestream}. 
This paradigm lets users specify outputs through examples rather than manual parameters.

\noindent\textbf{Attribute-based Training}
uses explicit labels, descriptors, or attribute degrees to specify target perceptual factors, such as speaker attributes, emotion, or speaking style. 
% 早期工作
Early methods manipulate voice attributes by disentangling linguistic content from extra-linguistic factors~\cite{benaroya2021beyond}.
% 更新的工作(情感编辑为例)
In recent works, taking emotion editing as an example, \cite{kreuk2022textless} use target emotion as the condition, while \cite{du2021disentanglement} disentangle emotional style from speaker identity to enable independent emotion transfer. Emo-CampNet~\cite{wang2024emotion} further incorporates emotion attributes into text-based speech editing to generate emotion-controlled edited regions.
% 总结一下
This paradigm supports controllable manipulation of interpretable perceptual factors.

\begin{figure*}[t]
    \centering
    \includegraphics[width=\textwidth]{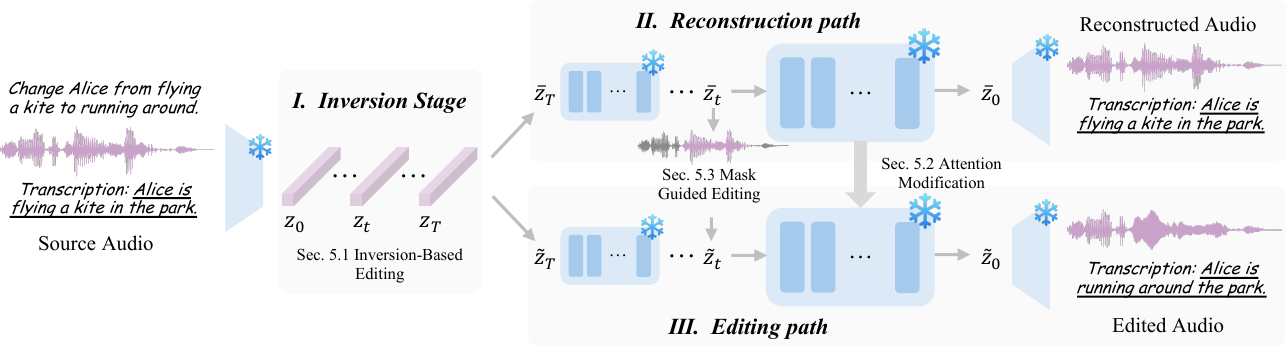}
    \caption{
Overview of training-free audio editing methods.
  }
    \label{fig:training-free}
    \vspace{-1em}
\end{figure*}

\subsection{Instructional Training}

% 介绍一下训练范式
Instructional training supervises audio editing models with $(\texttt{instruction}, \texttt{input}, \texttt{output})$ triplets, where natural-language requests specify edit operations, target regions, styles, or preservation requirements. 
% 对比过渡一下
Compared with reference or attribute based training, it provides more flexible and compositional control through natural language.

% 进展介绍
Representative works show a progression from template-based instruction following to more open-ended and compositional audio editing.
% 最开始的工作
\cite{wang2023audit} introduce instructional latent-diffusion editing for operations such as addition, removal, replacement, inpainting, and super-resolution. 
% 后续工作介绍
Later datasets and models, including SAO-Instruct~\cite{ungersbock2025sao} and MMEdit~\cite{tao2025mmedit}, extend this paradigm to free-form instructions and broader operations such as reordering and attribute modification. 
This framework has also expanded to speech and music, where instruction-following models support natural-language control over speech content, acoustic attributes~\cite{huang2024instructspeech,yan2025ming}, as well as music editing  and remixing~\cite{yan2025step,han2023instructme,zhang2024instruct}. 
Recent systems such as AudioChat~\cite{chen2026audiochat} further suggest a shift toward multi-task, multi-source, and interactive editing.

% 最后总结一下
By using language as a unified control interface, this paradigm moves audio editing from predefined task execution toward open-ended intent following in foundation-model-based editing.

\section{Training-Free Approaches}\label{sec:train-free}
% \input{latex/Tables/audio_edit_all_methods}

% 训练无关方法在不更新参数的情况下，将预训练音频生成模型适配到编辑任务中，主要通过操控推理阶段的反演、注意力、引导或掩码约束来实现。我们将现有方法归纳为三类常见范式，这些范式通常被组合使用，以提升编辑定位、源音频保持和可控性。由于基于 token 的自回归模型并不天然适合训练无关编辑，本节聚焦于非自回归范式，尤其是基于扩散模型的基础模型。
Training-free approaches adapt pretrained audio generative models to editing without parameter updates, mainly through inference-time inversion, attention, guidance, or mask constraints. 
As shown in Figure~\ref{fig:training-free}, we group them into three categories that are often combined to improve localization, preservation, and controllability. 
Since token-based autoregressive models are less suited to training-free editing, this section focuses on non-autoregressive paradigms, especially diffusion-based models.

\subsection{Inversion-Based Editing}
% 范式定义
Inversion-based editing maps source audio into the latent, noise, or trajectory space of a pretrained generative model, enabling edits by modifying conditions or sampling paths. It is particularly suited to diffusion and flow-matching models, where audio can be reconstructed through diffusion, latent, or ODE-based inversion~\cite{mokady2023null}.

% 代表工作
Pretrained latent diffusion models first show the potential of zero-shot audio operations, such as text-guided style transfer, inpainting, and super-resolution~\cite{liu2023audioldm}.
% DDPM
Inversion-based methods in DDPM further formulate training-free audio editing by inverting source audio into diffusion trajectories for text-guided or unsupervised editing~\cite{manor2024zero}. 
% 后续工作介绍
Subsequent methods improve preservation and editability by refining inversion, suppressing EOT artifacts, relaxing prompt constraints, or balancing source preservation with prompt adherence~\cite{jia2025audioeditor,paissan2023audio}.
% 一些工作拓展
This paradigm also extends to music and speech, where regularized, disentangled, or ODE-based inversion supports text-guided music editing, source-preserving recomposition, and controllable speech editing~\cite{lan2024high,liu2024medic,lv2026ast,tae2021editts}.

% 总结一下
Inversion-based methods reuse pretrained generators, but their performance depends on inversion accuracy, as reconstruction errors can weaken preservation or alter non-target regions.

\subsection{Attention Modification}
% 方法定义
Attention modification guides pretrained generative models by modifying or reusing internal attention patterns, often together with inversion, to control how textual conditions, audio events, and structural patterns are edited or preserved.

% 工作介绍 Cross-Attn & Self-Attn
Representative methods mainly rely on cross-attention control and self-attention preservation.
% Cross
Cross-attention methods utilize text-audio correspondences to localize or modify target events, using attention maps, word replacement, or prompt attention operations to reflect textual edits in latent or token representations~\cite{xu2024prompt,zhang2024musicmagus,sioros2025editgen}.
% self attention
Self-attention methods instead reuse source-side attention features to maintain musical or acoustic structure, with recent extensions to ReFlow~\cite{liu2022flow}, Diffusion Transformer~\cite{peebles2023scalable}, and localized temporal editing settings~\cite{liu2024medic,boudaghi2025musrec,jiang2025freeaudio}.

Attention modification improves localization, text-event alignment, and structural preservation, but depends on whether attention maps reflect perceptual events or musical structures.

\begin{table*}[ht]
    \centering
    \small
    \setlength{\tabcolsep}{8.5pt}
    \renewcommand{\arraystretch}{1.10}
    \caption{Overview of publicly available datasets related to audio editing. 
    Row colors indicate \protect\colorbox{speechrow}{speech}, \protect\colorbox{musicrow}{music}, and \protect\colorbox{audiorow}{general audio} datasets. 
    \textbf{Dur} denotes total hours, and \textbf{Diversity} denotes speakers, instruments, or event classes. 
    \textbf{Paired}, \textbf{Timestamp}, and \textbf{Caption} indicate (source, target) pairs, temporal alignment, and annotations, respectively.}
    \scalebox{0.83}{
    \begin{tabular}{lcccccc}
    \toprule
    \textbf{Name} & \textbf{Dur (hour)} & \textbf{Items} & \textbf{Diversity} & \textbf{Paired} & \textbf{Timestamp} & \textbf{Caption} \\
    \midrule

    \rowcolor{speechrow}
    LibriSpeech-Edit~\cite{lv2026ast} & 3.6 & 2000 & 40 & $\checkmark$ & \checkmark & $\checkmark$ \\
    \rowcolor{speechrow}
    Emilia~\cite{he2025emilia} & 101654 & -- & -- & $\times$ & \checkmark & $\checkmark$ \\
    \rowcolor{speechrow}
    WenetSpeech4TTS~\cite{ma2024wenetspeech4tts} & 12800 & 6283187 & -- & $\times$ & \checkmark & $\checkmark$ \\
    \rowcolor{speechrow}
    % DailyTalk~\cite{lee2023dailytalk} & 20 & 23507 & 2 & $\times$ & $\times$& $\checkmark$ \\
    \rowcolor{speechrow}
    SeniorTalk~\cite{chen2025seniortalk} & 55.53 & 101 & 202 & $\times$ & $\checkmark$ & $\checkmark$ \\
    \rowcolor{speechrow}
    AliMeeting~\cite{yu2022m2met} & 118.75 & 240 & 489 & $\times$ & $\checkmark$ & $\checkmark$ \\
    % \rowcolor{speechrow}
    % AMI~\cite{kraaij2005ami} & 100 & 171 & 187 & $\times$ & $\checkmark$ & $\checkmark$ \\

    \rowcolor{musicrow}
    Slakh2100~\cite{manilow2019cutting} & 145 & 2100 & 34 & $\checkmark$ & $\checkmark$ & $\checkmark$ \\
    \rowcolor{musicrow}
    MUSDB18~\cite{rafii2017musdb18} & $\sim$10 & 150 & 4 & $\checkmark$ & $\times$& $\times$ \\
    \rowcolor{musicrow}
    MAESTRO~\cite{hawthorne2018enabling} & 198.7 & 1276 & 1 & $\checkmark$ & $\checkmark$ & $\checkmark$ \\
    \rowcolor{musicrow}
    MusicCaps~\cite{agostinelli2023musiclm} & 15.34 & 5521 & 15 & $\times$ & $\times$& $\checkmark$ \\
    \rowcolor{musicrow}
    MTG-Jamendo~\cite{bogdanov2019mtg} & 3769.88 & 55609 & 20 & $\times$ & $\times$& $\times$ \\
    \rowcolor{musicrow}
    NSynth~\cite{engel2017neural} & 339.98 & 305979 & 1006 & $\times$ & $\checkmark$ & $\checkmark$ \\
    \rowcolor{musicrow}
    GTSinger~\cite{zhang2024gtsinger} & 80.59 & 1366 & 20 & $\checkmark$ & $\checkmark$ & $\times$ \\

    \rowcolor{audiorow}
    AudioSet~\cite{gemmeke2017audio} & 5789.78 & 2084320 & 527 & $\times$ & $\times$& $\times$ \\
    \rowcolor{audiorow}
    FSD50K~\cite{fonseca2021fsd50k} & 108.3 & 51197 & 200 & $\times$ & $\times$& $\times$ \\
    \rowcolor{audiorow}
    ESC-50~\cite{piczak2015esc} & 2.8 & 2000 & 50 & $\times$ & $\times$& $\times$ \\
    \rowcolor{audiorow}
    AudioCaps~\cite{kim2019audiocaps} & 127.8 & 46000 & 527 & $\times$ & $\times$& $\checkmark$ \\
    \rowcolor{audiorow}
    Clotho~\cite{drossos2020clotho} & $\sim$43.59 & 6974 & 7853 & $\times$ & $\times$& $\checkmark$ \\
    \rowcolor{audiorow}
    WavCaps~\cite{mei2024wavcaps} & 7565.2 & 403050 & -- & $\times$ & $\times$& $\checkmark$ \\
    \rowcolor{audiorow}
    % AudioSetCaps~\cite{bai2025audiosetcaps} & 5277.8 & 1900000 & 527 & $\times$ & $\times$& $\checkmark$ \\
    \rowcolor{audiorow}
    VoiceBank+DEMAND~\cite{valentini2017noisy} & $\sim$7.1 & 12396 & 30 & $\checkmark$ & $\times$& $\checkmark$ \\
    \rowcolor{audiorow}
    WHAM!~\cite{wichern2019wham} & 81.68 & 28000 & 117 & $\checkmark$ & $\times$& $\times$ \\
    \rowcolor{audiorow}
    % WHAMR!~\cite{maciejewski2020whamr} & 81.68 & 28000 & 117 & $\checkmark$ & $\times$& $\times$ \\
    \rowcolor{audiorow}
    DNS Challenge~\cite{dubey2024icassp} & 941.53 & 1560954 & 6360 & $\checkmark$ & $\times$& $\checkmark$ \\

    \bottomrule
    \end{tabular}}
    \vspace{-1em}
    \label{tab:audio_edit_datasets}
\end{table*}

\subsection{Mask Guided Editing}

% 方法定义
Mask guided editing specifies where to edit or preserve source audio through masks defined in waveform, latent, or source component spaces. 
This makes it suitable for localized editing, inpainting, restoration, and source-level manipulation.

Representative methods differ mainly in mask design. Explicit time frequency or segment-level masks support localized modification, inpainting, and restoration~\cite{liang2025audiomorphix,moliner2023diffusion}, while optimized masks enable source separation based manipulation~\cite{lee2025dgmo}. Latent channel and similarity guided masks further constrain which representations or regions are rewritten, supporting timbre transfer and music completion~\cite{lee2026diffusion,turland2025similarity}.

Mask guided methods offer explicit context or temporal control, but their performance depends on mask quality; inaccurate masks may cause boundary artifacts, incomplete edits, or unintended changes in preserved regions.

% \vspace{-0.5em}
\section{Resources}\label{sec:resources}

We review the resources supporting audio editing in the foundation model era, covering training data and evaluation protocols. And we further discuss scalable data construction tools in Appendix~\ref{app_sec:data_tool}.

\subsection{Open-Source Datasets}

Open-source audio datasets are essential for training and adapting audio foundation models. As summarized in Table~\ref{tab:audio_edit_datasets}, most are not direct editing datasets but provide transferable supervision from speech, singing, music, audio effects, and audio-text pairs for scalable task construction.

% \paragraph{Speech datasets}
% Speech resources mainly support content editing, speaker-preserving reconstruction, style control, and long-context speech modification. 
% LibriSpeech-Edit~\cite{lv2026ast} is one of the few datasets directly targeting speech editing, providing a public benchmark for localized text-based modification. 
% Emilia~\cite{he2025emilia} and WenetSpeech~\cite{ma2024wenetspeech4tts} are more suitable as high-quality speech resources for training speech generation backbones, with multi-speaker coverage and diverse recording conditions. 
% DailyTalk~\cite{lee2023dailytalk} provides colloquial multi-turn dialogues, making it useful for dialogue-style speech editing, while SeniorTalk~\cite{chen2025seniortalk} contributes speech from more diverse user groups. 
% For realistic long-context and multi-speaker scenarios, AMI~\cite{kraaij2005ami} and AliMeeting~\cite{yu2022m2met} provide meeting-style recordings that can support speaker-aware editing, dialogue continuation, and source-preservation evaluation.

\noindent\textbf{Speech Datasets}
Speech resources support content editing, speaker-preserving reconstruction, style control, and long-context speech modification. As shown in Table~\ref{tab:audio_edit_datasets}, they range from direct editing benchmarks such as LibriSpeech-Edit~\cite{lv2026ast}, to large-scale generation resources such as Emilia~\cite{he2025emilia} and WenetSpeech~\cite{ma2024wenetspeech4tts}, and dialogue or meeting corpora for conversational, multi-speaker, and long-context scenarios~\cite{lee2023dailytalk,chen2025seniortalk,kraaij2005ami,yu2022m2met}.

\noindent\textbf{Musical Datasets}
Music editing relies on resources with stems, symbolic structure, lyrics, tags, or semantic descriptions. As shown in Table~\ref{tab:audio_edit_datasets}, these resources cover stem and score-based manipulation~\cite{manilow2019cutting,rafii2017musdb18,hawthorne2018enabling}, semantic control through captions~\cite{agostinelli2023musiclm,bogdanov2019mtg}, and timbre, note-level, lyrics, melody, or expressive singing-style editing~\cite{zhang2024gtsinger}.

% \paragraph{General audio datasets}
% General audio resources mainly support sound-event editing, audio-language grounding, and denoising-oriented acoustic editing.
% % Sound Event
% AudioSet~\cite{gemmeke2017audio} provides broad coverage of sound events and acoustic scenes, and FSD50K~\cite{fonseca2021fsd50k} offers cleaner open-domain event annotations; 
% ESC-50~\cite{piczak2015esc} further provides a compact taxonomy for controlled event-level validation. 
% % Caption
% For audio-language supervision, AudioCaps~\cite{kim2019audiocaps} and Clotho~\cite{drossos2020clotho} pair audio clips with natural-language captions, while WavCaps~\cite{mei2024wavcaps} and AudioSetCaps~\cite{bai2025audiosetcaps} scale this direction to larger caption resources. 
% % Denoising
% For denoising and restoration, VoiceBank+DEMAND~\cite{valentini2017noisy} provides noisy-clean speech pairs, WHAM!~\cite{wichern2019wham} and WHAMR!~\cite{maciejewski2020whamr} introduce noisy and reverberant mixtures, and DNS Challenge~\cite{dubey2024icassp} covers large-scale real-noise scenarios.

\noindent\textbf{General audio datasets}
General audio resources support sound-event editing, audio-language grounding, and acoustic restoration. As shown in Table~\ref{tab:audio_edit_datasets}, they cover event and scene annotations~\cite{gemmeke2017audio,fonseca2021fsd50k,piczak2015esc}, caption-based audio-language supervision~\cite{kim2019audiocaps,drossos2020clotho,mei2024wavcaps,bai2025audiosetcaps}, and noisy, reverberant, or real-noise mixtures for denoising and restoration~\cite{wichern2019wham,dubey2024icassp}.

% % 最后的总结
% Despite these resources, large-scale datasets tailored to instruction-based audio editing remain scarce. 
% % 具体展开当下的现状
% Most datasets provide raw audio, labels, captions, stems, or restoration pairs, but rarely combine input audio, edit instructions, target regions, and edited outputs. 
% Current systems therefore rely on synthetic pairs, task-specific benchmarks, or resources transferred from speech synthesis, source separation, audio captioning, and music generation. 
% Standardized and instruction-aligned editing datasets remain a key bottleneck for training and evaluating general-purpose audio editing models.

Despite these resources, large-scale instruction-based audio editing datasets remain scarce. Existing datasets provide only partial supervision and rarely pair input audio, edit instructions, target regions, and edited outputs in a unified format. Thus, current systems still depend on synthetic pairs or adapted resources, while standardized instruction-aligned datasets remain a key bottleneck.

\subsection{Evaluation Protocols and Metrics}

Compared with audio generation, audio editing evaluation must assess both edit success and preservation of non-target regions, in addition to output quality. We sum up existing protocols along four dimensions: edit success and instruction adherence, preservation and locality, temporal and structural consistency, and audio quality and naturalness.

% \noindent\textbf{Instruction Adherence}
% Edit success measures whether the intended modification has been correctly applied. 
% In speech editing, this evaluation should cover both linguistic correctness and attribute-control correctness. For content-level edits, ASR-based WER/CER~\cite{cao2023comparative} directly measures the mismatch between the edited speech and the target text, while for attribute-level edits, emotion2vec\cite{ma2024emotion2vec} is able to provide speech emotion representations that allows emotion-oriented evaluation, making it useful for assessing whether the intended emotional attribute has been achieved.
% For instruction-following audio editing, CLAP-based audio-text similarity~\cite{elizalde2023clap} can serve as a semantic proxy for whether the edited audio matches the textual instruction, especially for event- or attribute-level changes. 
% In music editing, edit success is usually decomposed into task-specific sub-goals, such as tag prediction accuracy, event classification accuracy, pitch accuracy, or rhythm-related errors, rather than being reduced to a single scalar metric~\cite{copet2023simple}.

\noindent\textbf{Instruction Adherence}
Instruction adherence evaluates whether the edited audio follows the intended textual or task-specific requirement. 
% In speech editing, it covers both linguistic correctness and attribute-control accuracy: ASR-based WER/CER~\cite{cao2023comparative} measures mismatch with the target text, while emotion representations from emotion2vec~\cite{ma2024emotion2vec} support emotion-oriented evaluation. 
% For audio editing, CLAP-based audio-text similarity~\cite{elizalde2023clap} serves as a semantic proxy for matching textual instructions, especially for event- or attribute-level changes. 
% In music editing, adherence is often decomposed into task-specific goals, such as tag prediction, event classification, pitch accuracy, or rhythm errors~\cite{copet2023simple}.
In speech editing, it includes linguistic correctness and attribute-control accuracy: ASR-based WER/CER~\cite{cao2023comparative} quantifies mismatch with the target transcript, whereas emotion representations from emotion2vec~\cite{ma2024emotion2vec} support affective evaluation. 
For general audio, CLAP-based audio-text similarity~\cite{elizalde2023clap} is widely used as a semantic proxy for instruction matching. 
Music editing typically decomposes adherence into task-specific criteria, including tag prediction, event classification, pitch accuracy, and rhythm errors~\cite{copet2023simple}.

\noindent\textbf{Preservation and Locality}
Preservation and locality quantify whether content outside the intended edit scope remains unchanged. 
Speaker identity is commonly assessed with speaker-embedding cosine similarity~\cite{snyder2018x}, while non-target acoustics can be compared at the waveform, spectrogram, or embedding level~\cite{wang2020complex,ragano2024nomad}. 
Restoration-oriented tasks often adopt reference-based metrics such as PESQ~\cite{rix2001perceptual}, STOI~\cite{taal2010short}, and SI-SDR~\cite{le2019sdr} to evaluate perceptual quality, intelligibility, and distortion relative to clean references. 
More generally, locality-aware evaluation should penalize both under- and over-editing within the specified time span, speaker, event, or source component.

\noindent\textbf{Temporal and structural consistency}
% Temporal and structural consistency measures whether the edited output remains coherent in time and preserves relevant structural patterns. 
% For speech editing, boundary accuracy can be evaluated by the mean or median error between manual and automatic boundaries~\cite{williams2024analysis}, while WDTW~\cite{lv2026ast} can measure local temporal fidelity between edited and target utterances at the word level.
% For music editing, structure-aware metrics are particularly important: melody accuracy~\cite{wu2024music} measures whether the generated melody follows the target melodic contour, Rhythm F1~\cite{lan2024musicongen} evaluates rhythmic alignment, and dynamics correlation~\cite{sioros2025editgen} quantifies the consistency between the  loudness curve and generated dynamic changes. 
Temporal and structural consistency examines whether the edited signal remains coherent over time and preserves relevant organization. 
Boundary accuracy is commonly quantified by the mean or median error between manual and automatic boundaries in speech editing~\cite{williams2024analysis}; WDTW~\cite{lv2026ast} further measures word-level temporal fidelity between edited and target utterances.
For music editing, structure-aware metrics are particularly important: melody accuracy~\cite{wu2024music} evaluates whether the output follows the target contour, Rhythm F1~\cite{lan2024musicongen} measures rhythmic alignment, and dynamics correlation~\cite{sioros2025editgen} captures consistency between loudness trajectories and generated dynamics. 
These metrics are useful for edits that must keep prosody, rhythm, melody, or long-range arrangement.

\noindent\textbf{Audio quality and naturalness}
Audio quality and naturalness evaluate whether the edited audio sounds realistic, continuous, and free from obvious artifacts or synthetic discontinuities.
MOS listening tests~\cite{rec2006p} remain the most direct protocol, while MOSNet~\cite{lo2019mosnet}, DNSMOS~\cite{reddy2021dnsmos}, and NISQA~\cite{mittag2021nisqa} provide scalable automatic estimates.
Recent edit-aware evaluators, including AuditScore, AuditEval~\cite{jia2025towards}, TTA-Bench~\cite{wang2026tta}, AudioEval~\cite{wang2025audioeval}, T2A-Feedback~\cite{wang2025t2a}, and MLLM-based judges~\cite{pu2025judge}, better connect perceptual quality with instruction relevance and source fidelity.
In music, FAD~\cite{kilgour2018fr} estimates distributional distance to reference audio, while MuseCPBench~\cite{vishe2025musecpbench} targets structural preservation.
% Because automatic metrics are still proxies, human evaluation remains important for subjective style control, instruction following, and complex music or soundscape editing.
Given the limits of automatic proxies, human evaluation remains essential for subjective style control, instruction following, and complex audio edits.

\section{Conclusion}

This survey provides a structured review of audio editing in the foundation model era, connecting task formulation with foundational architectures, adaptation strategies, and supporting resources. 
Our analysis shows that foundation models have greatly improved editing controllability and flexibility, while future research can further advance precise local editing, source preservation, disentangled control, and reliable evaluation.
We hope this survey can support future research toward more general and practical audio editing systems.

% \clearpage

\section*{Limitations}

This survey focuses on audio editing in the era of foundation models, where recent progress is largely driven by large-scale generative backbones, audio-language models, codec language models, diffusion models, and flow-matching frameworks. \textbf{Consequently, we mainly review foundation-model-based audio editing methods}. Earlier paradigms, including statistical parametric methods, signal-processing-based workflows, and early task-specific deep learning approaches, are discussed only as background rather than surveyed in detail. In addition, this survey primarily considers monaural audio editing. Spatial audio editing, such as stereo, binaural, or multi-channel Ambisonics-based editing, involves specialized issues of spatial rendering, room acoustics, and perceptual localization, and is therefore beyond the main scope of this review.

\section*{Ethical Considerations}

Although this survey itself raises no immediate ethical concerns, the reviewed audio editing methods may introduce risks when applied in practice. We highlight two aspects. (1) \textbf{Data licensing and consent.} Users should respect dataset licenses and obtain proper permission before collecting or repurposing web-hosted speech, music, or sound recordings, especially when they involve personal voices, singing performances, copyrighted music, or private acoustic scenes. (2) \textbf{Misuse of editing models.} Foundation-model-based audio editing can realistically alter speech content, speaker identity, emotion, background sounds, and music. Without safeguards, it may enable voice impersonation, unauthorized modification, copyright infringement, or deceptive media generation. Practitioners should follow usage policies and regulations, while future work should explore provenance tracking, watermarking, and edited-audio detection.

% \section*{Acknowledgements}

% This work was supported by National Natural Science Foundation of China under Grant No.U24A20326.

\bibliography{custom}

\clearpage
\appendix
\section{Scope}
In this survey, we focus on works that make direct contributions to audio editing in the era of foundation models. To ensure a precise and focused discussion, we adopt two main inclusion criteria: (1) the task should center on audio editing, which we define as modifying the acoustic attributes, instances, or content of an existing audio recording, without transformations so substantial that they amount to generating an entirely new audio sample; (2) the method should rely on mainstream audio foundation model paradigms.
Accordingly, we do not cover works primarily focused on audio generation, nor do we provide an extensive discussion of signal-processing-based audio editing methods. In addition, to maintain a focused scope, spatial audio\footnote{Spatial audio refers to multi-channel audio formats, such as binaural stereo and first-order Ambisonics (FOA).} and related editing techniques are beyond the main scope of this survey.

\section{Discussion}
\subsection{Foundation Models with Discrete and Continuous Representations}
The surveyed methods suggest that audio editing with foundation models is primarily shaped by the underlying audio representation and generation process. Codec language models discretize audio into token sequences, which provides a compact interface for infilling, continuation, and generation conditioned on editing tasks. This makes them particularly effective for speech editing, where linguistic content, speaker identity, and local temporal structure can be explicitly modeled. By contrast, diffusion and flow matching models operate on continuous acoustic representations, such as spectrograms or latent features, and are therefore more suitable for high fidelity reconstruction, fine grained acoustic manipulation, and generation constrained to target regions.

These architectural choices lead to different failure modes. Discrete token models may lose spectral details due to codec compression and can suffer from autoregressive latency or error propagation, especially in music and general audio with complex timbral dynamics. Continuous generative models better preserve local acoustic structure, but their editing behavior is highly sensitive to target localization, source conditioning, mask design, and preservation objectives. In both cases, the key technical difficulty is not only to generate plausible target audio, but also to enforce edit locality, so that the modified region satisfies the editing condition while the surrounding content remains perceptually unchanged. Overall, progress in audio editing will likely depend on architectures that jointly support semantic controllability, continuous high fidelity synthesis, and explicit preservation of regions outside the edit target, rather than optimizing generation quality alone

\subsection{Training-based and Training-free Editing Paradigms}
Training-based and training-free methods reflect different assumptions about how editing ability is acquired. Training-based methods learn editing behaviors from paired data, pseudo pairs, or instruction style supervision. Their advantage lies in stability: when the task and data distribution are clearly defined, they can directly optimize edit success, preservation of non-edited regions, and output naturalness. This makes them suitable for reliable instruction following, attribute control, and repeated use of fixed editing functions. However, their performance depends heavily on the coverage and quality of training data. Given the scarcity of large-scale instruction aligned audio editing datasets, many methods still rely on synthetic pairs, task specific pipelines, or adapted resources, which may introduce distribution bias and limit generalization to open domain requests.

Training-free methods instead reuse pretrained generative models without parameter updates. Through inversion, attention modification, or mask constraints, they exploit the editing capacity already encoded in foundation models. This paradigm avoids costly editing specific training and can be applied to new prompts, domains, or edit types, especially when paired editing data are unavailable. Its main limitation is weaker controllability and less predictable preservation. Editing quality often depends on accurate inversion, meaningful attention localization, and reliable separation between edited and preserved regions.

The two paradigms are complementary. Training-based methods are better suited to robust systems under well specified task settings, while training-free methods offer flexibility for open domain and low resource editing. Future audio editing systems may combine training-based adaptation for instruction following and preservation with training-free mechanisms for localization, source control, and test time refinement.

\section{Detailed Description of the Task}
\label{app_sec:task}
We note that the proposed taxonomy does not define mutually exclusive task boundaries. In practice, audio editing operations are often compositional, and a single instruction may belong to multiple task classes. We therefore view the taxonomy as a functional organization based on the primary edited information layer, rather than a strict one-to-one mapping between an instruction and a category.

A typical example is the instruction "remove the background crowd noise from the speech". This operation can be viewed as denoising or restoration when the crowd sound is treated as an unwanted degradation. It can also be regarded as instance-level deletion when the crowd is modeled as a separable sound source in the audio scene. This example shows that real-world audio editing tasks are often multi-label by nature. In the following descriptions, we assign each task to the category that best reflects its dominant editing goal, while allowing overlaps across acoustic, semantic, and instance-level editing.

\subsection{Acoustic Editing}
% 响度编辑
\noindent\textbf{Loudness and volume editing} constitute one of the most common forms of acoustic editing. This capability reorganizes auditory attention by making certain components more prominent or by adjusting the balance between foreground and background sounds~\cite{jiang2024listen,choi2021amss,liang2024wavcraft}. For example, in advertising scenarios, adjusting key segments and the foreground-background balance can largely enhance expressiveness and listener engagement.

% 音频去噪
\noindent\textbf{Denoising and restoration} are another important class of acoustic editing, focusing on the suppression of undesired degradations, such as background noise, codec artifacts, and local corruption, while preserving useful acoustic cues and source characteristics~\cite{liu2022voicefixer}. Rather than erasing all environmental traces, these tasks aim to remove unwanted degradations while preserving the naturalness of the original audio~\cite{fu2021metricgan+}.

% 混响编辑
\noindent\textbf{Reverberation editing} modifies room-acoustic attributes, such as room response, early reflections, and reverberation tails~\cite{chen2022visual}. It adjusts spatial and temporal decay patterns to change the perceived acoustic environment while preserving the underlying source. Related tasks such as dereverberation and room-acoustic control suggest that room impulse responses can serve as editable acoustic conditions rather than merely distortions to be removed~\cite{bahrman2024speech}.

% 声音均衡
\noindent\textbf{Audio equalization} is an essential manifestation of acoustic editing that traditionally requires substantial domain expertise. It modifies the spectral energy distribution of the raw audio to control perceptual qualities such as brightness, warmth, sharpness, or muffledness. Also, recent studies represent that equalization can be learned and parameterized by neural networks, which enables descriptive language control, such as ``brighter'', ``warmer'', or ``more distant''~\cite{venkatesh2022word}.

\subsection{Semantic Editing}

% 内容编辑/语言学编辑
\noindent\textbf{Linguistic editing} revises explicit language-bearing content. In speech, it inserts, deletes, or replaces words and phrases while preserving speaker timbre, surrounding acoustics, and smooth transitions across edited regions~\cite{tan2021editspeech}. In music, related operations include modifying lyrics or short vocal phrases while maintaining melody, singer identity, and accompaniment consistency~\cite{zhang2022editsinger}. Such capabilities support correction, updating, and localized revision without regenerating the entire recording.

\noindent\textbf{Expressive editing} changes how audio is delivered or emotionally perceived. For speech, it covers emotion~\cite{wang2024emotion}, prosody, speaking rate, pitch contours, accent, and paralinguistic cues such as whispers, laughter, sighs, and hesitation~\cite{kim2025fillerspeech}. For general audio, it may adjust the affective character of a sound event, making it more tense, gentle, dramatic, or playful. These tasks require controlled transformation of expressive attributes while preserving content, source identity, and naturalness in edited audio.

\noindent\textbf{Stylistic editing} targets high-level musical and stylistic properties, such as genre, mood, melodic expression, rhythm, or arrangement-level characteristics~\cite{zhang2024musicmagus}. Beyond music, stylistic editing may also change the presentation style of speech or general audio, such as making speech more formal, colloquial, or advertisement-like, or making a sound more realistic~\cite{anastassiou2024voiceshop,jin2023voice}. Since these attributes are often subjective and language-described, this type of editing requires models to align style descriptions with coherent audio transformations.

\subsection{Instance Editing}
\noindent\textbf{Instance replacement} substitutes a target entity with another instance while preserving the surrounding context. In speech, this often corresponds to voice conversion, where the speaker identity is changed while the linguistic content and local acoustics are retained~\cite{wang2021vqmivc,qian2019autovc}. In general audio, it may involve replacing one sound event with another~\cite{xu2024prompt}; in music, it may replace a vocal, instrumental, or local musical component~\cite{han2023instructme}. The main challenge is to ensure that the substituted instance remains coherent with the acoustic context, temporal alignment, and source relationships of the original recording~\cite{fu2025object}.

\noindent\textbf{Instance deletion and extraction} remove or isolate a target entity from a mixture. Deletion suppresses the target while preserving the remaining scene~\cite{ellis2025recomposer}, whereas extraction isolates it for separate use or further editing~\cite{liu2024separate}. Typical examples include target-speaker extraction~\cite{vzmolikova2019speakerbeam}, sound-event removal~\cite{fu2025object}, and music stem separation~\cite{defossez2021hybrid}, all requiring accurate instance discrimination and preservation of non-target component within complex mixtures.

\noindent\textbf{Instance insertion} introduces a new entity into an existing recording, such as a speaker turn, sound event, instrument stem, or vocal phrase. Common examples include inserting sound effects into an acoustic scene~\cite{wang2023audit} or extending a dialogue with an additional utterance~\cite{jin2017voco}. Unlike generation from scratch, insertion requires the added instance to be coherent with the original acoustic environment, temporal layout, and, when applicable, musical arrangement~\cite{rouard2025musicgen,peng2024voicecraft}.

\noindent\textbf{Instance overlay} layers or emphasizes an instance without fully replacing existing components. It covers foreground enhancement, soundscape remixing, background ambience enrichment, and music-layer addition~\cite{ellis2025recomposer}. It may enhance a target speaker in a noisy conversation, add crowd ambience to an outdoor scene, or layer a rhythmic component over existing music. Unlike insertion, overlay editing preserves the original mixture while adjusting the salience, density, or texture of selected instances, making it useful for enriching or rebalancing an audio scene under global coherence constraints~\cite{strano2025stage}.

\section{Data Construction Tools}\label{app_sec:data_tool}

Given the scarcity of complete instruction-based editing datasets, data construction tools are essential for converting raw audio resources into reliable, structured, and reusable editable supervision. We group these tools by the artifacts they produce: temporal boundaries for locating edit regions, semantic annotations for describing edit targets, and paired examples for defining desired edited outputs.

\paragraph{Temporal localization tools}
Temporal localization tools determine where an edit should be applied. 
In speech editing, this is often achieved through text-audio alignment: Praat~\cite{boersma2021praat} supports manual boundary inspection and acoustic analysis, MFA~\cite{mcauliffe2017montreal} provides automatic phone- or word-level forced alignment, and WhisperX~\cite{bain2023whisperx} generates word-level timestamps for long-form speech. When the editable unit is defined by speaker activity rather than text, pyannote~\cite{bredin2023pyannote} and VAD tools~\cite{karan2024transformer} can provide speaker-active segments. 
For general audio, sound event detection models ~\cite{kong2020panns,li2023ast} produce event-level activity boundaries. % 在原来cite基础上再找一个
In music editing, temporal localization often relies on pitch or note-level cues: Parselmouth~\cite{jadoul2018introducing}, RMVPE~\cite{wei2023rmvpe}, and CREPE~\cite{kim2018crepe} extract time-varying F0 contours, while ROSYOT~\cite{li2024robust} and MusicYOLO~\cite{wang2022musicyolo} provide precise note-level onset and offset annotations for editing in structured workflows.

\paragraph{Semantic annotation tools}
After editable areas are localized, semantic annotation tools assign interpretable descriptions or labels to them. For speech, ASR and alignment systems~\cite{gao2023funasr,radford2023robust} provide transcripts that can serve as content annotations.
For non-speech audio, tagging, detection, and captioning tools can produce sound-event labels, scene labels, or natural-language descriptions~\cite{chen2022hts}; SELD models~\cite{adavanne2018sound,guirguis2021seld}, for example, provide event-oriented annotations with localization information. 
Attribute annotations are also important for expressive editing: emotion2vec~\cite{ma2024emotion2vec} offers speech emotion representations that can support large-scale affective labeling. 
Recent audio-language models, such as Qwen3-Omni~\cite{xu2025qwen3}, can further assist caption generation and tag-to-description conversion, making raw labels more suitable for instruction-style supervision.

\paragraph{Pair construction tools}
% 这里缺少一段voice conversion的内容 补一句话在speech editing后面哈
Pair construction tools define the desired edited output by creating input-target examples. For speech editing, controllable TTS systems such as MaskGCT~\cite{wang2024maskgct} and StyleTTS~\cite{li2025styletts} can synthesize paired examples with controlled content, speaking style, or prosody. Voice conversion systems such as AutoVC~\cite{qian2019autovc} and
YourTTS~\cite{casanova2022yourtts} can transform speaker identity while preserving linguistic content, providing paired examples for speaker replacement or timbre-level editing.
For acoustic editing, degradation simulation can construct restoration pairs by adding noise, reverberation, bandwidth limitation, or codec artifacts to clean recordings. 
For instance-level editing, source separation and remixing tools provide another route: Open-Unmix~\cite{stoter2019open}, Spleeter~\cite{hennequin2020spleeter}, and Demucs~\cite{rouard2023hybrid} decompose music into stems, while AudioSep~\cite{liu2024separate} and SAM-Audio~\cite{shi2025sam} extend separation to open-domain or prompt-based scenarios with text, visual, or temporal-span prompts. 
These tools enable the construction of insertion, deletion, extraction, replacement, and overlay pairs from existing audio mixtures for supervised editing.

\section{Future Directions and Challenges}

Despite the success achieved in audio editing in the foundation model era, there are still limits to address in future work across practical applications.

\subsection{Complex Editing}

Although recent audio editing methods have already achieved promising progress in specific domains, their editing capability remains limited when facing complex and entangled audio scenes. 
% 举例说明问题
Existing audio editing methods are still largely shaped by domain-specific task definitions. 
Speech editing mainly focuses on text replacement, speaker identity preservation, and speech naturalness~\cite{peng2024voicecraft,huang2024instructspeech}. 
General audio editing is often formulated as event-level addition, deletion, or replacement~\cite{wang2023audit}. 
Music editing further requires the preservation of harmony, rhythm, instrumentation, and long-range musical structure~\cite{han2023instructme}. 
Although these task-specific designs have enabled effective editing in constrained settings, they remain far from a unified editing framework that can precisely modify arbitrary audio components while preserving task-irrelevant content in open scenarios.

% 说明实际的挑战
The central challenge stems from the highly coupled nature of audio. Real-world recordings often entangle semantic events, speaker identity, acoustic texture, background ambience, rhythm, spatial cues, and reverberation within a single waveform. As a result, audio editing must not only generate plausible target content, but also localize the editable region, modify the intended attributes, and preserve non-target context in timbre, timing, loudness, and acoustic coherence. Such requirements become especially difficult for overlapping sources, long-form speech, dense soundscapes, and music, where object boundaries are often weak or implicit.

% 未来方向
Future research should move toward disentangled and object-aware audio editing. One possible direction is to develop source-aware representations that separate editable objects from non-target contexts, enabling more reliable local generation and preservation. Another direction is to introduce structured control signals, such as temporal regions, event boundaries, speaker labels, source masks, acoustic attributes, or spatial trajectories, to reduce the ambiguity of free-form instructions. Recent unified audio generation models~\cite{yang2023uniaudio,valle2025fugatto} suggest the potential of handling speech, sound, singing, and transformation within a general framework. However, future audio editing systems should go beyond unified generation and clearly support object localization, attribute-level modification, and non-target preservation across speech, music, and general audio.

\subsection{Robustness Under Open-Domain}
% 跑出问题
Robustness under open-domain and real-world conditions remains a central challenge for audio editing. 
Existing methods often perform well on predefined tasks, clean inputs, or carefully constructed instructions, but their behavior becomes unstable when the audio contains noise, reverberation, overlapping sources, long-range dependencies, or ambiguous user intents. These failures arise from two coupled difficulties: the model must ground the editing instruction to the correct acoustic target, and it must preserve non-target content under complex acoustic conditions in real-world mixtures.

% 当前技术不足和面临的挑战
Different modeling paradigms expose this challenge in different forms. 
% codec为什么不太行
Codec language model-based methods formulate editing as token infilling, contextual continuation, or text-conditioned generation, and have shown strong potential in speech editing, as demonstrated by VoiceCraft~\cite{peng2024voicecraft} and SpeechX~\cite{wang2024speechx}. 
However, their effectiveness depends heavily on stable text-acoustic alignment and high-quality tokenization. When applied to music or general audio, weak semantic boundaries, dense source mixtures, and long-term structural constraints may lead to token-level errors, context drift, and fidelity degradation. 
% diffusion为什么不太行
Diffusion and flow-matching models, by contrast, are strong at generating high-fidelity acoustic details in continuous latent spaces, making them suitable for editing timbre, background, sound texture, and musical details. 
Yet their local controllability remains limited: instruction-guided latent diffusion editing has shown promising results~\cite{wang2023audit,gao2026rfm}, but local modifications may affect non-target regions, especially in music and complex soundscapes where acoustic components are strongly correlated~\cite{han2023instructme}.

% 未来可能的研究方向
Future work should therefore move beyond improving generation quality alone and focus on robust editing behavior. 
One direction is to improve instruction grounding, enabling models to infer the intended target, editable region, desired attribute, and preservation scope from natural language. 
Another direction is to strengthen long-context modeling and preservation mechanisms, so that edits remain stable in long-form or multi-source audio with extended dependencies.
Iterative refinement, self-verification, retrieval-augmented editing, and multi-stage correction may further help models recover from uncertain or failed edits, especially in multi-turn editing scenarios.

\subsection{Faithful and Specific Evaluation}

Faithful evaluation remains a critical bottleneck for audio editing. 
% 第一层挑战，评测不够对口
The first challenge lies in the mismatch between existing benchmarks and editing tasks. 
Current audio resources mainly focus on generation, tagging, captioning, understanding, or audio-language reasoning, rather than editing behavior itself. 
Datasets such as AudioSet~\cite{gemmeke2017audio} and AudioCaps~\cite{kim2019audiocaps} support audio tagging, captioning, and audio-language alignment, but lack the $(\texttt{instruction}, \texttt{input}, \texttt{output})$ triplets needed to verify whether a specific edit is correctly performed. Similarly, benchmarks such as MARBLE~\cite{yuan2023marble} and AudioBench~\cite{wang2025audiobench} assess music understanding or general audio-language understanding and reasoning, but are not designed to measure edit success, non-target preservation, or instruction-following quality in edited outputs.
As a result, current protocols often conflate generation quality with editing quality.

% 挑战二，评测不够准
The second challenge comes from the multi-dimensional nature of audio editing.
Unlike audio generation, editing requires modifying only the intended content while preserving the remaining audio and maintaining perceptual coherence.
A faithful protocol should therefore jointly assess target modification, non-target preservation, acoustic naturalness, and instruction following~\cite{wang2023audit}.
These dimensions are related but not equivalent: an edited sample may sound natural yet fail the requested edit, or follow the instruction while unintentionally changing speaker identity, background ambience, rhythm, or acoustic texture.

% 挑战3: 工具不够多
The third challenge is the limitation of existing evaluation tools.
Most automatic metrics capture only partial aspects of editing quality. Text-audio similarity reflects semantic alignment but not preservation~\cite{elizalde2023clap};
ASR-based accuracy and speaker similarity are useful for speech editing but less applicable to music or general audio~\cite{morris2004and,snyder2018x}; 
acoustic quality predictors can detect artifacts but cannot judge whether the edit is correct~\cite{reddy2021dnsmos}. 
Recent audio-language and multimodal models~\cite{tian2025step,xu2025qwen3,comanici2025gemini} offer a promising direction for automatic judges, but they must be calibrated against human listening results to avoid overestimating instruction alignment while missing subtle artifacts, temporal discontinuities, or unintended non-target changes in realistic evaluations.

Future evaluation should shift from generation-centered to editing-centered assessment. 
Reliable benchmarks should provide explicit annotations of target regions, edit operations, preservation regions, and control signals, enabling separate measurement of edit success and non-target preservation. 
Metrics should also be multi-dimensional, covering instruction following, target modification, content preservation, acoustic naturalness, and perceptual coherence. Human listening tests remain essential for calibrating automatic evaluators. 
Although human-aligned audio editing judges that combine semantic understanding with perceptual sensitivity offer a scalable direction, their reliability must be validated across diverse tasks, domains, and acoustic conditions. Such benchmarks and evaluators are crucial for advancing audio editing toward robust real-world applications at practical scale.

\end{document}